\documentclass[aps,prd,twocolumn,preprintnumbers,superscriptaddress,floatfix]{revtex4}

\usepackage{multirow, graphicx,amssymb,url,mathrsfs,amsmath}
\usepackage{eucal,wrapfig,boxedminipage,setspace,subfigure}
\usepackage{amsxtra,amstext,latexsym,dsfont,amsfonts}
\usepackage{color,eucal}
\usepackage[colorlinks,hyperindex]{hyperref} 

       \def\d  {\delta}

\def\IR{{\hbox{{\rm I}\kern-.2em\hbox{\rm R}}}}
\def\IB{{\hbox{{\rm I}\kern-.2em\hbox{\rm B}}}}
\def\IN{{\hbox{{\rm I}\kern-.2em\hbox{\rm N}}}}
\def\IC{\,\,{\hbox{{\rm I}\kern-.59em\hbox{\bf C}}}}
\def\IZ{{\hbox{{\rm Z}\kern-.4em\hbox{\rm Z}}}}
\def\IP{{\hbox{{\rm I}\kern-.2em\hbox{\rm P}}}}
\def\IH{{\hbox{{\rm I}\kern-.4em\hbox{\rm H}}}}
\def\ID{{\hbox{{\rm I}\kern-.2em\hbox{\rm D}}}}

\newcommand{\beq}{\begin{equation}}
\newcommand{\eeq}{\end{equation}}
\newcommand{\bea}{\begin{eqnarray}}
\newcommand{\eea}{\end{eqnarray}}

\begin{document}

\voffset 1cm

\newcommand\sect[1]{\emph{#1}---}

\title{Holographic Nambu Jona-Lasinio Interactions}

\author{Nick Evans}
\affiliation{STAG Research Centre \&  Physics and Astronomy, University of
Southampton, Southampton, SO17 1BJ, UK}

\author{Keun-Young Kim}
\affiliation{School of Physics and Chemistry, \\
Gwangju Institute of Science and Technology, Gwangju 500-712, Korea }

\begin{abstract}
NJL interactions are introduced into the D3/ probe D7 system using Witten's double trace operator prescription which includes the operator as a classical term in the effective potential. In the supersymmetric system they do not induce chiral symmetry breaking which we attribute to the flat effective potential with quark mass in the supersymmetric theory.  If additional supersymmetry breaking is introduced then standard NJL behaviour is realized. In examples where chiral symmetry breaking is not preferred such as with a B field plus an IR cut off chiral condensation is triggered by the NJL interaction at a second order transition after a finite critical coupling.  If the model already contains chiral symmetry breaking, for example in the B field case with no IR cut off, then the NJL interaction enhances the quark mass at all values of the NJL coupling. We also consider the system at finite temperature: the temperature discourages condensation but when combined with a magnetic field we find regions of parameter space where the NJL interaction triggers a first order chiral transition above a critical coupling.

\end{abstract}

\maketitle

\newpage

The Nambu Jona-Lasinio (NJL) model \cite{Nambu:1961tp} is a long standing phenomenological model of dynamical chiral symmetry breaking. In this paper we will seek to reproduce its results in a holographic setting \cite{Maldacena:1997re}. We choose the simple D3/probe D7 model of ${\cal N}=2$ quark multiplets in a background of large $N_c$ ${\cal N}=4$ super Yang-Mills theory \cite{Karch:2002sh}. We will show that if an additional source of supersymmetry breaking is present in addition to the NJL four fermion operator then the holographic models indeed reproduce the traditional dynamics expected. The key understanding is that the effective potential of the model consists of two pieces: the first is the effective potential of the theory without an NJL term provided in holography by the bulk action, the second is a UV surface term representing the NJL operator's presence. In the basic D3/ probe D7 model the bulk contribution is zero due to the supersymmetry of the system and the NJL surface term does not generate chiral condensation. The addition of an IR relevant perturbation that breaks the supersymmetry leads to a bulk contribution to the effective potential that falls at large $m$ and NJL behaviour is then restored. We display this with a magnetic field perturbation.  We will also explore examples where the NJL interactions enhances an IR condensation mechanism, and the effects of temperature. We find an example with both temperature and magnetic field present that generates a first order transition triggered by an NJL interaction.
\vspace{-0.75cm}

\section{\bf NJL Model} 

The NJL model \cite{Nambu:1961tp} consists of massless quarks plus a four fermion interaction term	
\begin{equation} \label{ffo} \Delta {\cal L} = {g^2 \over \Lambda_{UV}^2} \bar{q}_L q_R \bar{q}_R q_L + h.c \,. \end{equation}
The theory generates the famous gap equation
\begin{equation} 1 = {g^2 \over 4 \pi^2} \left( 1 - {m^2 \over  \Lambda_{UV}^2} \log \left[(\Lambda_{UV}^2 + m^2)/ m^2\right] \right) \,, \end{equation}
which has non-zero solutions for $m$ when $g^2 > 4 \pi^2$.

 \begin{figure}[]
 \centering
   {\includegraphics[width=6.5cm]{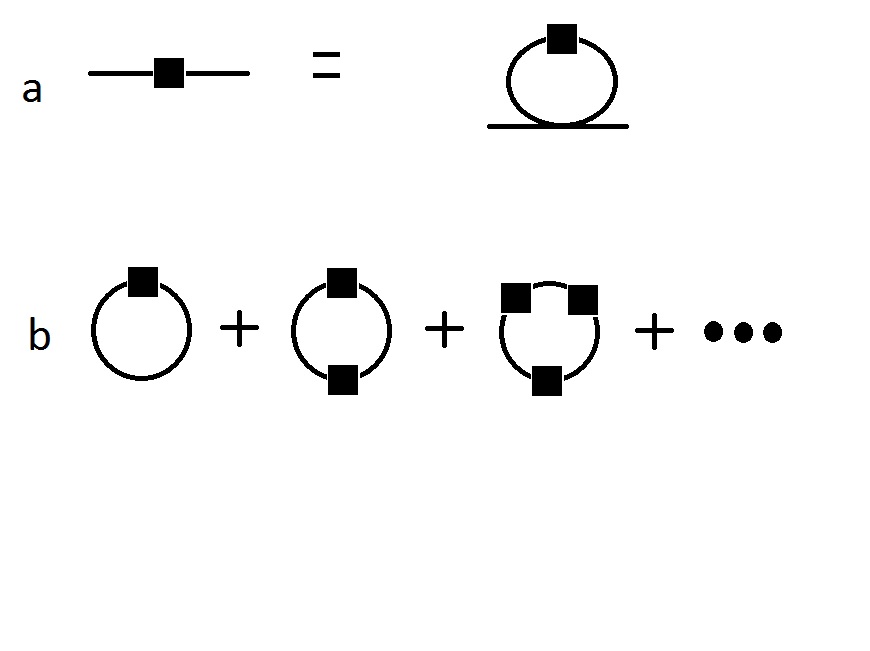} \label{}} \vspace{-1cm}
  \caption{ The NJL model gap equation (a) and the loop diagrams (b) contributing the effective potential of the model. The box represents a mass insertion. }
             \label{NJL1}
\end{figure}

\begin{figure}[]
 \centering
   {\includegraphics[width=6.5cm]{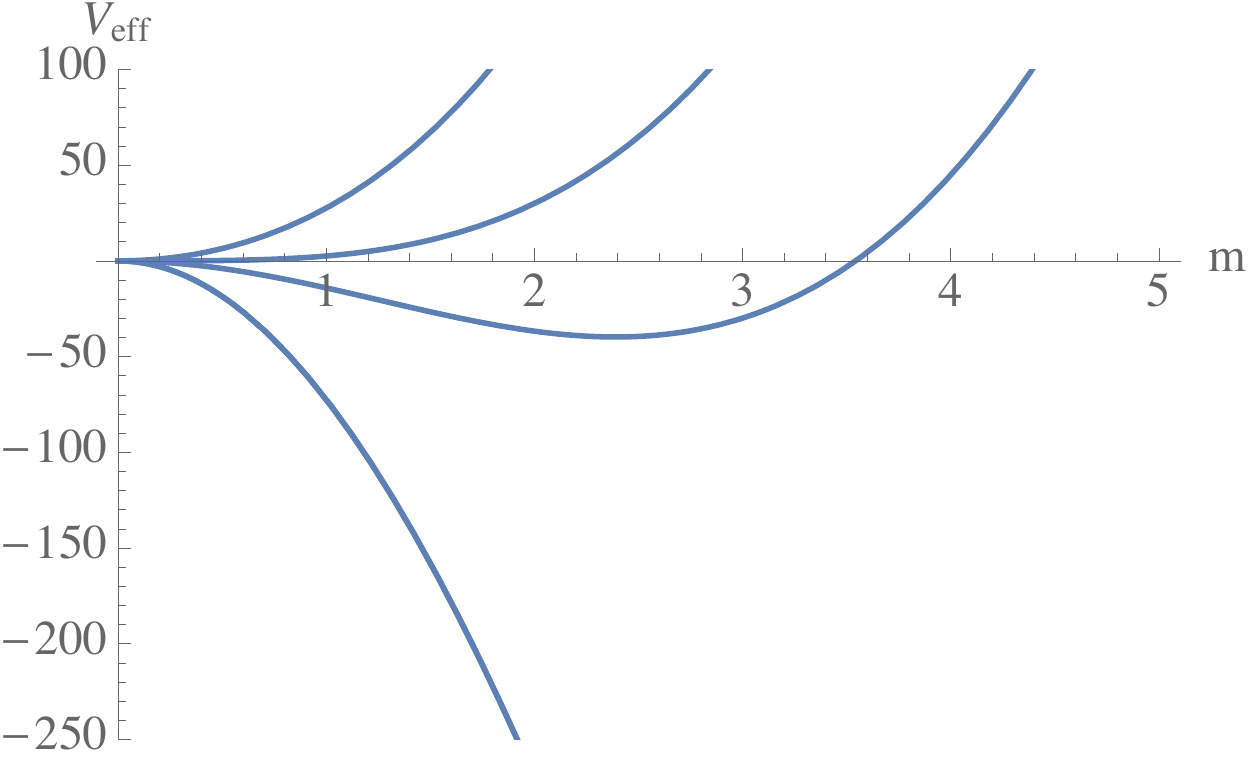} \label{}} 
  \caption{ The form of the effective potential against quark mass in the NJL model for a range of $g$. At infinite $g$ the loop contributions of Fig \ref{NJL1} generate an unbounded potential. At finite $g> g_c$ a non-zero $m$ minimizes the potential. For $g<g_c$ $m=0$ is the minimum. The transition is second order at $g_c$. }
            \label{NJL2}
\end{figure}

We can understand the origin of this gap equation in two ways. Firstly directly from the diagrammatic  Schwinger Dyson equation shown in Fig \ref{NJL1}(a) and secondly by minimizing the effective potential resulting from the diagrams in Fig \ref{NJL1}(b) plus the classical four fermion operator. Note that this represents a mean field approximation, justifiable at large $N$. The effective potential \cite{Coleman:1973jx} from the massive quark loops is given by
\begin{equation} 
\Delta V_{\rm eff} =  - \int_0^{\Lambda_{UV}} {d^4k \over (2 \pi)^4} \mathrm{Tr} \log (k^2 + m^2)  \,.
\end{equation}	
Without the NJL term the quark mass is a parameter of the theory and one should not minimize this potential with respect to $m$. In the presence of the NJL interaction we add
\begin{equation} 
V_{\rm eff} = {g^2 \over \Lambda^2_{UV}} \left(\langle \bar{q}_L q_R \rangle\right)^2 = {m^2 \Lambda_{UV}^2\over g^2} \,, \label{vterm} 
\end{equation}
where we have used the classical result in the presence of a quark condensate
$m = {g^2 \over \Lambda_{UV}^2}\langle \bar{q}_L q_R \rangle$.
Now since $g$ might induce $m$ we do seek to minimize the potential against $m$. We plot the form of the potential in Fig \ref{NJL2} for various values of $g$. For infinite $g$ the loop term dominates and the potential is unbounded preferring $m \rightarrow \infty$. As $g$ decreases from infinity the four fermion term enters making the potential bounded at large $m$. For a regime of $g$ down to the critical coupling there is a non-zero $m$ at the potential minimum. Below the critical $g$ $m=0$ is the minimum. There is therefore a second order transition to chiral symmetry breaking at $g^2=g_c^2= 4 \pi^2$. 
 
\section{\bf D3/D7 System}

The simplest  realization of quarks in holography is provided by placing probe D7 branes in the AdS$_5 \times S^5$ geometry representing the inclusion of quenched ${\cal N}=2$ quark multiplets  in ${\cal N}=4$ super Yang Mills 
\cite{Karch:2002sh}. We write the metric as
\begin{equation} ds^2 = r^2 dx_{3+1}^2 + {1 \over r^2} \left( d \rho^2 + \rho^2 d \Omega_3^2 + dL^2 + L^2 d \varphi^2 \right) \,, \end{equation}
where $\rho^2 + L^2 = r^2$ with $r$ the AdS radius.
The probe, which fills the $x_{3+1}, \rho, \Omega_3$  directions at constant $\varphi$  has action 
\begin{equation} \label{action}  S = - {\cal{A}}  \int d \rho \ {\cal L} \,,  \hspace{1cm}   {\cal L} = e^{-\phi} \rho^3 \sqrt{1 + L^{'2}} \,, 
\end{equation}

where  ${\cal{A}}$ is the D7 brane tension times the volume factor of the space $x_{3+1}$ and  $\Omega_3$. In AdS the dilaton, $\phi$, is a constant.
 Varying the action gives
\begin{equation}  \delta S = 0 = \int d \rho \left(\partial _\rho{\partial {\cal L }\over \partial L'}  - {\partial {\cal L }
\over \partial L} \right)  \delta L  + \left. {\partial {\cal L }\over \partial L'} \delta L \right|_{{UV, IR}} \,.  \end{equation}
Since the action only depends on $L'$ there is a conserved quantity $2c$ from which we learn
\begin{equation} L' = { - 2 c \over \sqrt{\rho^6- 4 c^2}} \label{conserved} \,, \end{equation}
and hence at large $\rho$ 
\begin{equation}  L = m + {c \over \rho^2} \,. \end{equation}
The only regular solutions over all $\rho$ have $c=0$ as can be seen from (\ref{conserved}) and hence are the constant embeddings $L=m$. $m$ is interpreted as proportional to the hypermultiplet mass. Since $m$ is a parameter of the theory we insist on it being fixed in the UV i.e. we pick the UV boundary condition $\delta L=0$ for the Euler Lagrange problem. In the IR we pick $ {\partial {\cal L} \over \partial L'}  = {\rho^3 L' \over \sqrt{1 + L^{'2}}}=0 $ which is satisfied since $L'=0$ for the $L=m$ solution.

\section{\bf Double Trace Operator}

Witten's prescription \cite{Witten:2001ua} to include double trace operators is to set a new appropriate UV boundary condition. Here we wish to include a term like that in (\ref{ffo}) which at a classical level implies the quark mass is given by $g^2 \langle \bar{q}_L q_R \rangle/ \Lambda_{UV}^2$ so, if we treat the parameter $c$ as the condensate then we want at the UV boundary
\begin{equation} \label{boundary} c = {\Lambda_{UV}^2 m \over g^2} \,.  \end{equation}
We can achieve this by adding a UV boundary action term
\begin{equation} \Delta S _{UV}=  {\cal A} {L^2 \Lambda_{UV}^2 \over g^2 + 1 } \label{bound} \,.  \end{equation}
Now at the UV boundary we no longer require after variation of $L$ $\delta L=0$ but allow $L$ to change and instead impose 
\begin{equation} 0 = {\partial {\cal L} \over \partial L'}  + {2 L \Lambda_{UV}^2 \over g^2 + 1}  \,,  \label{cond1} \end{equation}
which gives the required $c,m$ relation at leading order for large $\Lambda_{UV}$.

We maintain the IR boundary condition $L'=0$. 

In the base supersymmetric theory we have already found the solutions of the Euler Lagrange equations that satisfy this IR condition - they are just $L=m$. These though only satisfy the new UV boundary term when $m=0$ since $c$ is always zero. The four fermion interaction does not therefore behave in a standard NJL model fashion here - it does not generate a mass no matter how large the coupling. 

This may seem surprising but is best understood by considering the effective potential. First note that the boundary term \eqref{bound} we have added to the action is  up to a multiplicative constant,  basically the classical effective potential term that is added in the NJL model (\ref{vterm}) since in the UV $L=m$ and at large $\Lambda_{UV}$:  ${g^2 + 1 \over \Lambda_{UV}^2} \sim {g^2 \over \Lambda_{UV}^2}$. If we begin at $g \rightarrow \infty$ then the extra UV boundary term vanishes and the effective potential for $m$ is just the original action of the D7 probe evaluated on the solutions $L=m$. All these solutions have the same potential since the action (\ref{action}) only depends on $L'$. This analysis mimics that of the NJL model but here there is a sharp contrast to the NJL model where an unbounded potential with $m$ is found in this limit. In the supersymmetric case presumably cancellations in vacuum loops between the quarks and squarks remove any $m$ dependence. At finite $g$ the bulk action continues not to discriminate between different $m$ solutions but the UV term simply chooses the lowest value of $m$ which is zero.

It is clear though how we should progress to reconjure NJL-like dynamics; if we can raise the action of solutions with small $m$ relative to those with large $m$ in the base theory the expected NJL dynamics will reappear. In fact this is easy to do by including any IR relevant supersymmetry breaking perturbation. Such an interaction will generate some non-zero $L'$ in (\ref{action}) raising the action for embeddings associated with small quark masses, sensitive to the IR, but large mass solutions will asymptote to the supersymmetric theory's lowest possible action.  As a first example we will consider the case where the perturbation is  a $U(1)$ baryon number magnetic field.

\section{\bf Magnetic Field Perturbation}

\begin{figure}[]
 \centering 
{\includegraphics[width=6.5cm]{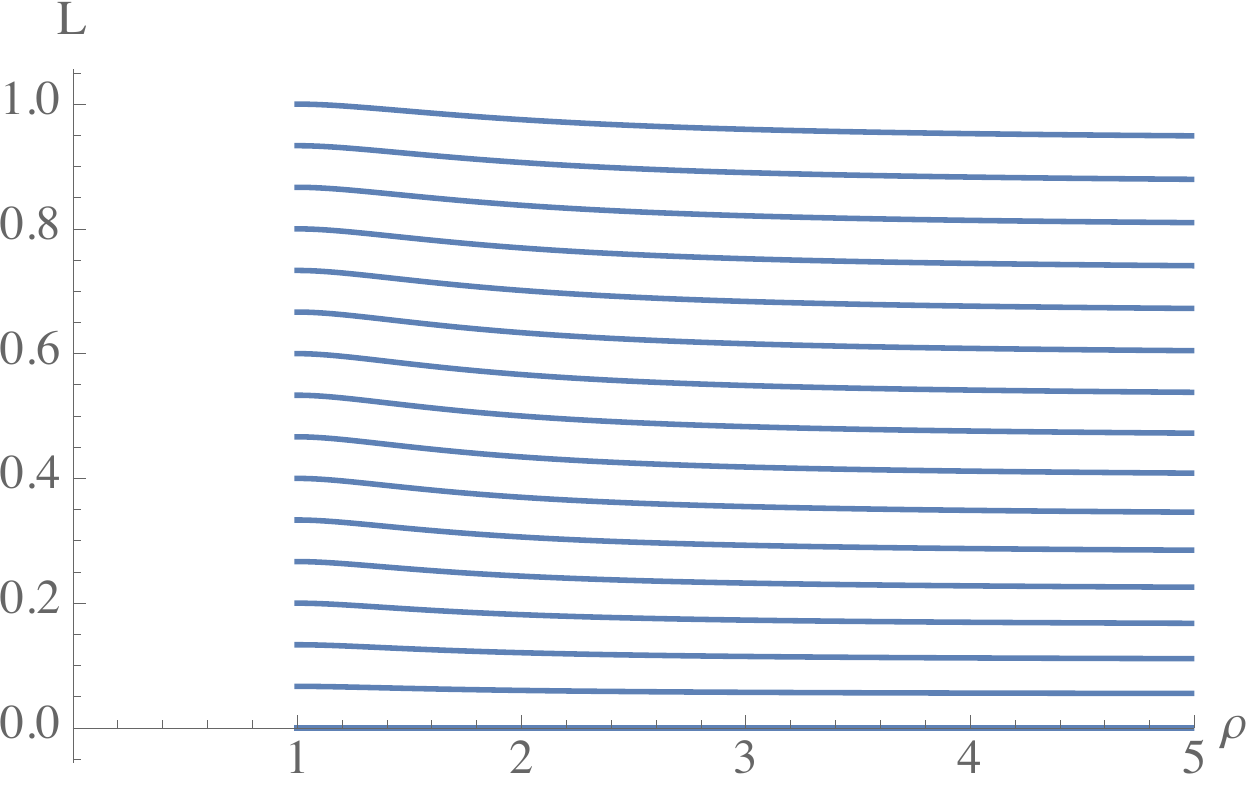}  \vspace{0.5cm} \\
\includegraphics[width=6.5cm]{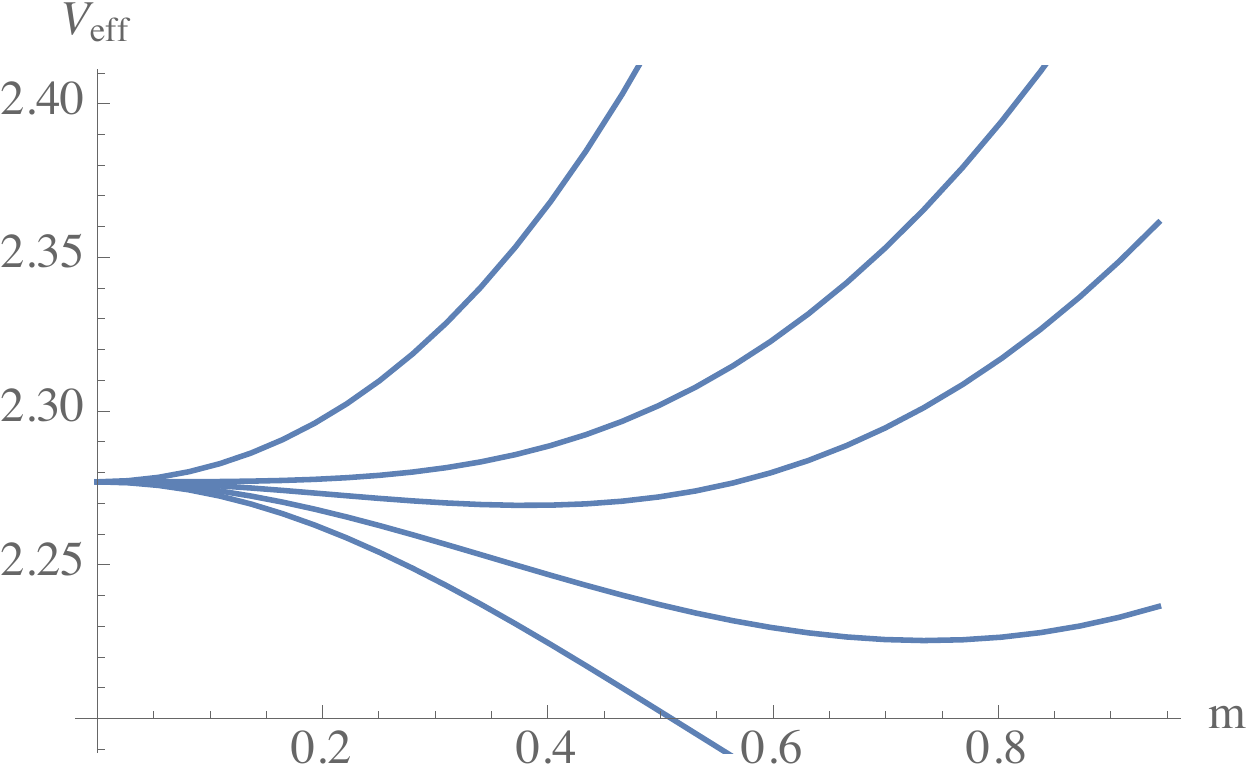} \vspace{0.5cm} \\
\includegraphics[width=6.5cm]{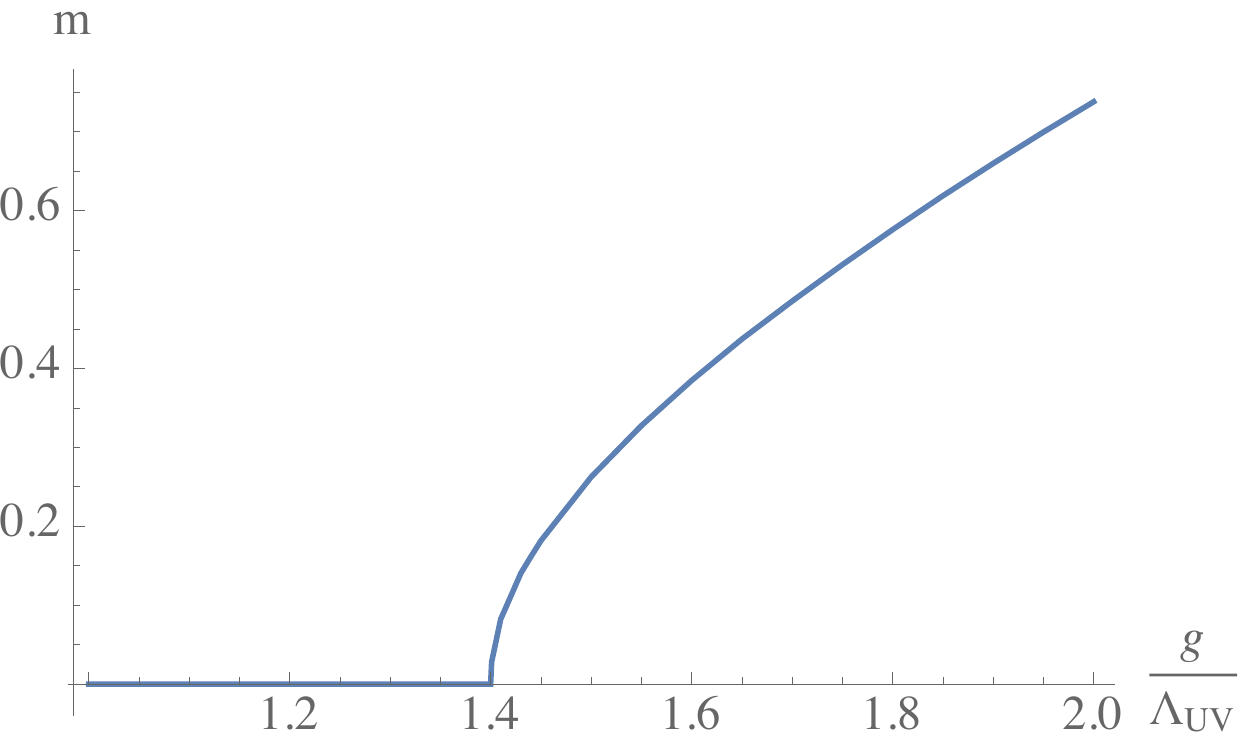} }
  \caption{ The D3/D7 model with a magnetic field, IR cut off and NJL interaction at a UV cut off. \\
 {\bf Top:} are the IR boundary condition preserving embeddings with $B=1$ and $\Lambda_{IR}=1$.  
\\ {\bf Middle:} is the potential including UV surface term for those embeddings with $g/\Lambda_{UV}= 1, 1.4,1.6,,2,3$ from top to bottom. Here $\Lambda_{UV}=100$.
\\{\bf Bottom:} $m$ that minimizes the potential versus $g/\Lambda_{UV}$ showing the second order transition at a critical value of $g$. }
            \label{fig:3}
\end{figure}

A $U(1)$ baryon number magnetic field, $B$, can be introduced into the D3/probe D7 model through the world volume vector of the D7 brane \cite{Filev:2007gb}. $B$ enters in the action  \eqref{action} as an effective dilaton term
\begin{equation} e^{-\phi} = \sqrt{ 1 + {B^2 \over (\rho^2 + L^2)^2}} \end{equation}
The magnetic field by itself induces chiral symmetry breaking. Since we are interested primarily in the chiral symmetry breaking properties of the NJL interaction term it is convenient initially to turn off the deep IR effects of $B$. This is easily done by setting an IR cut off on the model $\Lambda_{IR} > B^{1/2}$. The $B$ field then enters into the bulk action to discriminate between different $m$ solutions but does not have sufficient IR power to drive chiral symmetry breaking itself (we will also study the $\Lambda_{IR} \rightarrow 0$ case shortly). 

Without the NJL operator one would solve the Euler Lagrange equations subject in the IR to $L'(\Lambda_{IR})=0$. The solutions run to different UV values of $L \simeq m$  and each gives the appropriate solution for that value of $m$. In Fig \ref{fig:3} (top) we plot examples of such solutions (with $B^{1/2}=\Lambda_{IR}=1$). 

Now we can introduce the NJL operator by changing the UV boundary condition. For each choice of $g$ we could search amongst the solutions of Fig \ref{fig:3} (top) for those that satisfy the new UV boundary condition (\ref{boundary}). Alternatively we can compute their action as an effective potential from (\ref{action}) and (\ref{bound}) (note that (\ref{bound}) still holds since $e^{-\phi} \rightarrow 1$ in the UV) - the solution we seek will be the minimum. In fact the action is UV divergent generating a term  $\Lambda_{UV}^4$. Since we will have an explicit UV cut off we could just leave this large piece in the results but we choose to show plots where we subtract a term $\Lambda_{UV}^4$  from all values we quote.  We compute both the integral over the Lagrangian evaluated on the solution in the bulk plus the boundary term (\ref{bound}). The results for some different $g$ and for $\Lambda_{UV}=100$ are shown in the middle plot of Fig  \ref{fig:3}. 

The model displays all the characteristics of the NJL model. The effective potential from the integration of the action over the bulk falls with larger mass in the absence of the NJL interaction.

When the NJL interaction surface term is included the potential turns up and becomes bounded. There is a second order phase transition with the NJL interaction at a critical value as we show in the bottom graph of Fig  \ref{fig:3} which shows the growth of the mass against $g$.

\section{\bf Magnetic Field Induced Symmetry Breaking Enhanced By NJL}

\begin{figure}[]
 \centering 
{\includegraphics[width=6.5cm]{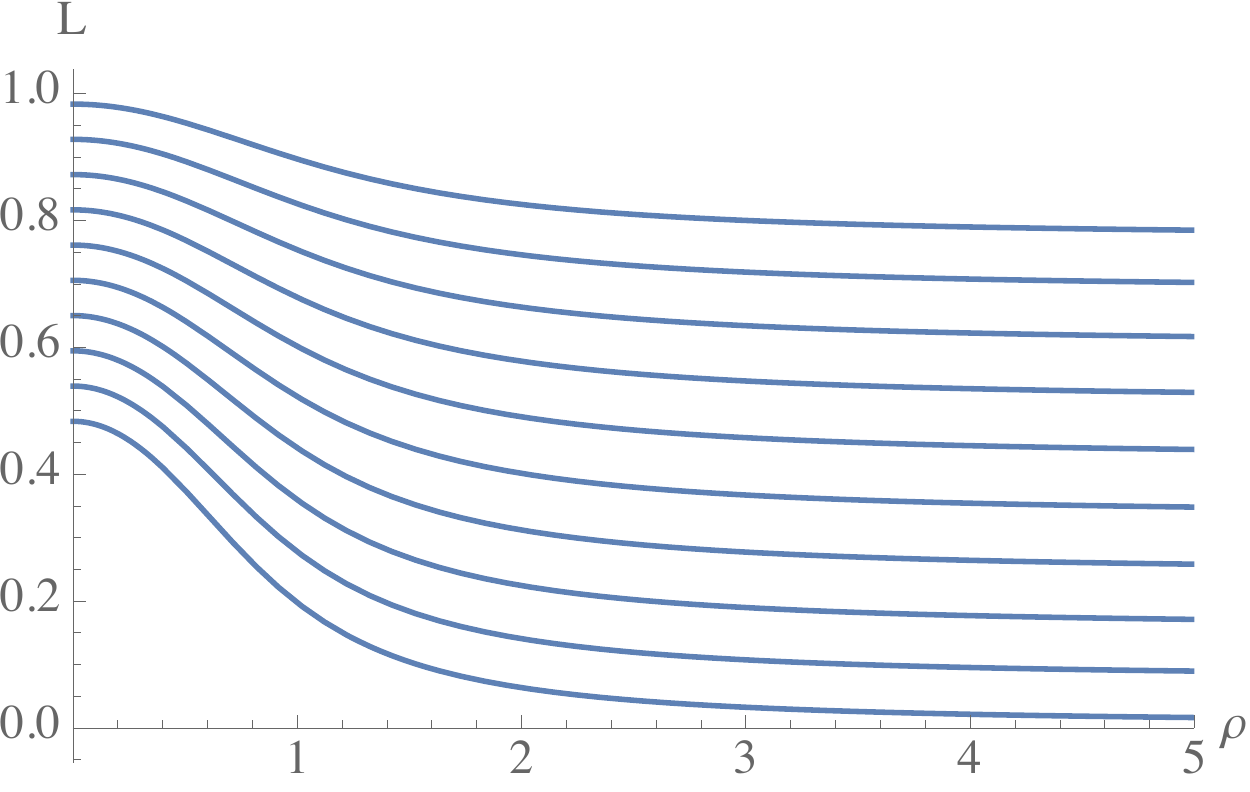}  \vspace{0.5cm} \\
\includegraphics[width=6.5cm]{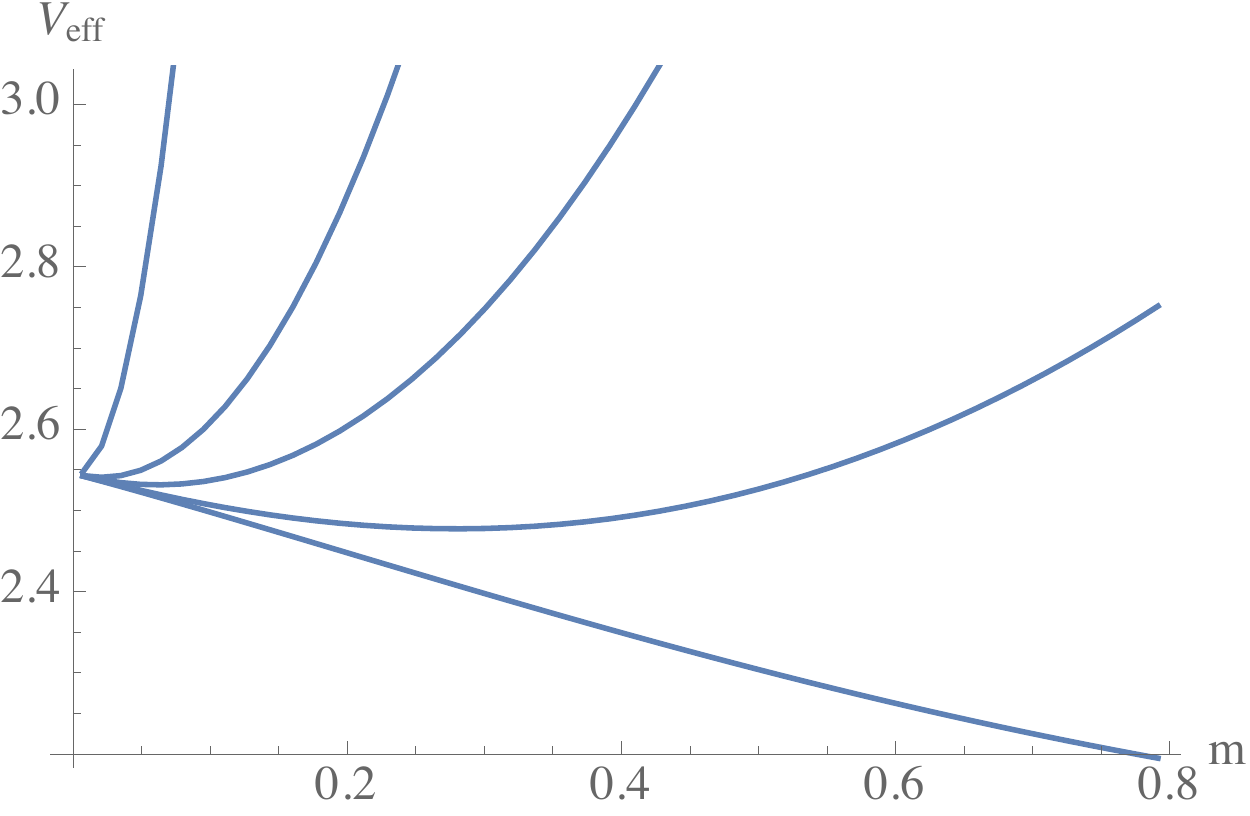} \vspace{0.5cm} \\
\includegraphics[width=6.5cm]{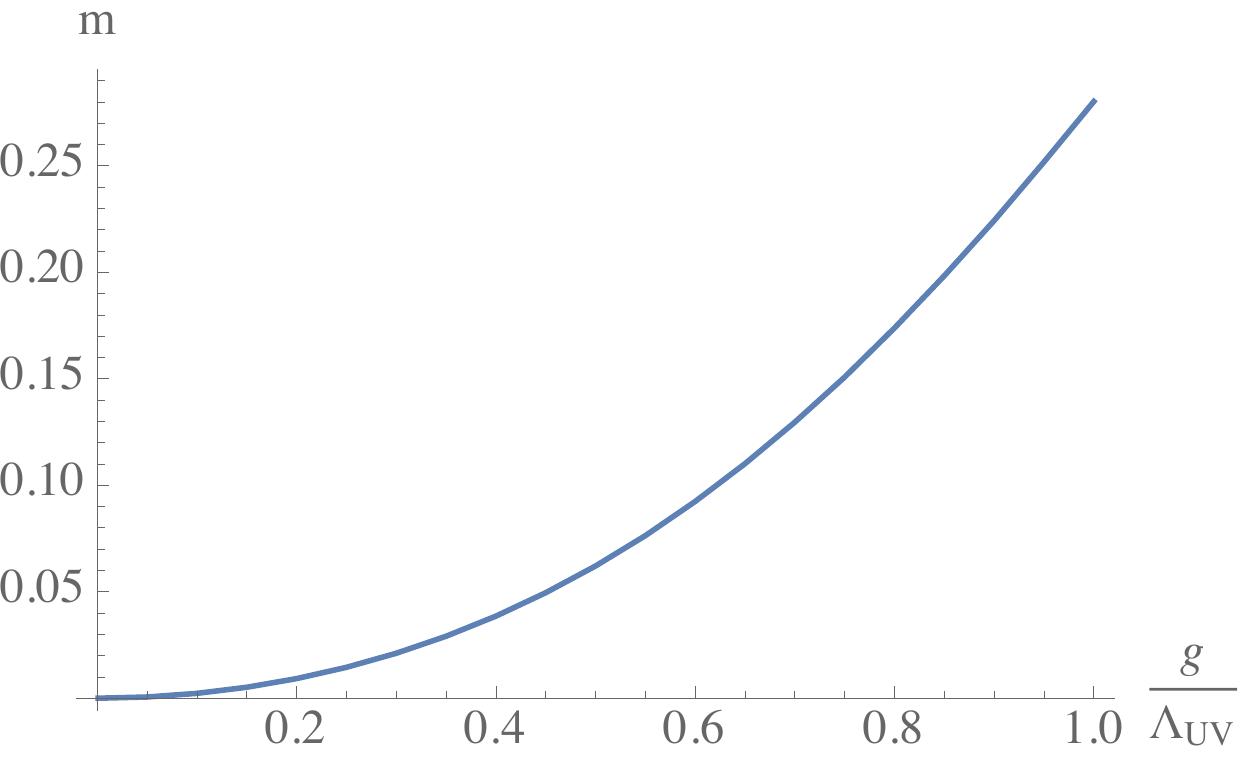} }
  \caption{ The D3/D7 model with a magnetic field, no IR cut off, and NJL interaction at a UV cut off.\\
 {\bf Top:} are the IR boundary condition preserving embeddings with B=1 and $\Lambda_{IR}=0$. 
\\ {\bf Middle:} is the potential including UV surface term for those embeddings with $g/\Lambda_{UV}= 0.1,0.3,0.5,1,3$ from top to bottom. Here $\Lambda_{UV}=100$.
\\{\bf Bottom:}$m$ that minimizes the potential versus $g/\Lambda_{UV}$ showing the smooth switch on of the mass as $g$ rises from zero. }
            \label{fig:4}
\end{figure}

Next we can study the theory with the magnetic field in the limit where the IR cut off is taken to zero. Now the B field itself introduces chiral symmetry breaking and the NJL interaction will act to enhance it. The top figure of Fig \ref{fig:4} shows the chiral symmetry breaking induced by the B field on its own - here we plot regular embeddings without the NJL interaction i.e. those with $L'(0)=0$ for the case $B=1$. The embedding for $m=0$ in the UV displays curvature now and bends away from the origin - there is a non-zero condensate $c$ asymptotically and the IR clearly breaks the U(1)$_\varphi$ symmetry. 

In the central figure of Fig \ref{fig:4} we show the effective potential against $m$ with a UV cut off of 100 in units of $B^{1/2}$. When the NJL term is absent the potential is again unbounded but for finite NJL coupling the effective potential is minimized for a non-zero value of $m$.

 Here, as the lower figure in Fig \ref{fig:4}  shows, the NJL operator generates a mass no matter how small the NJL operator is (this is because there is already a condensate present due to $B$). As the NJL operator coupling is increased the mass grows. This is consistent with gap equation analysis of chiral symmetry breaking theories enhanced by NJL operators (see for example \cite{King:1990hd}). 

\section{\bf Finite Temperature}

Another possible IR deformation is finite temperature \cite{Babington:2003vm,Mateos:2006nu}. In appropriate coordinates for embedding a D7 brane the AdS-Schwarzschild metric with the horizon at $r=r_H$ is \cite{Evans:2011mu}
\begin{equation} 
\begin{split}
ds^2 = &w^2 \left(-g_t dt^2 + g_x dx_3^2\right)  \\
&+ {1 \over w^2} \left(d \rho^2 +\rho^2 \d \Omega_3^2 + dL^2 + L^2 d \varphi^2\right)  \,,
\end{split}
\end{equation}
where $\sqrt{2}w = \sqrt{r^2 +\sqrt{r^4 - r_H^4}} $  (note  $w=r$ in the  UV and $\sqrt{2}w_H \equiv r_H$ ) and
\begin{equation} g_t = {(w^4 - w_H^4)^2 \over  w^4(w^4+w_H^4)}, \hspace{1cm} g_x = {w^4 + w_H^4 \over  w^4} \,. \label{met1} \end{equation}
The DBI action for the D7 is \eqref{action} with an effective dilaton term
\begin{equation}  \label{dilat}
\begin{split}
& e^{-\phi}  =   \\ 
&  \sqrt{\left(1 - {w_H^8 \over (\rho^2 + L^2)^4} \right)^2  
 + \frac{B^2}{(\rho^2 + L^2)^2}\left(1 - {w_H^4 \over (\rho^2 + L^2)^2} \right)^2      } \   \,. 
\end{split}
\end{equation}
Note here we have also included a magnetic field. 

\begin{figure}[]
 \centering 
{\includegraphics[width=6.5cm]{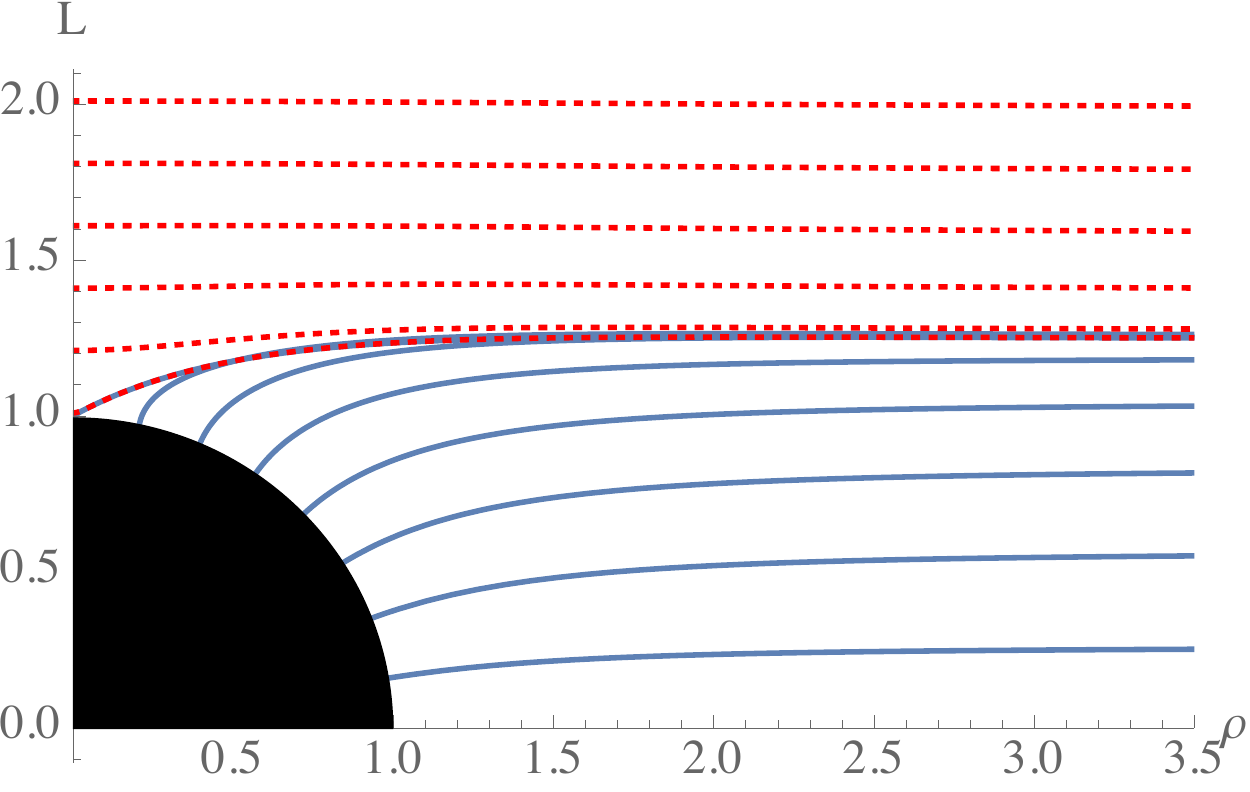}  \vspace{0.5cm} \\
\includegraphics[width=6.5cm]{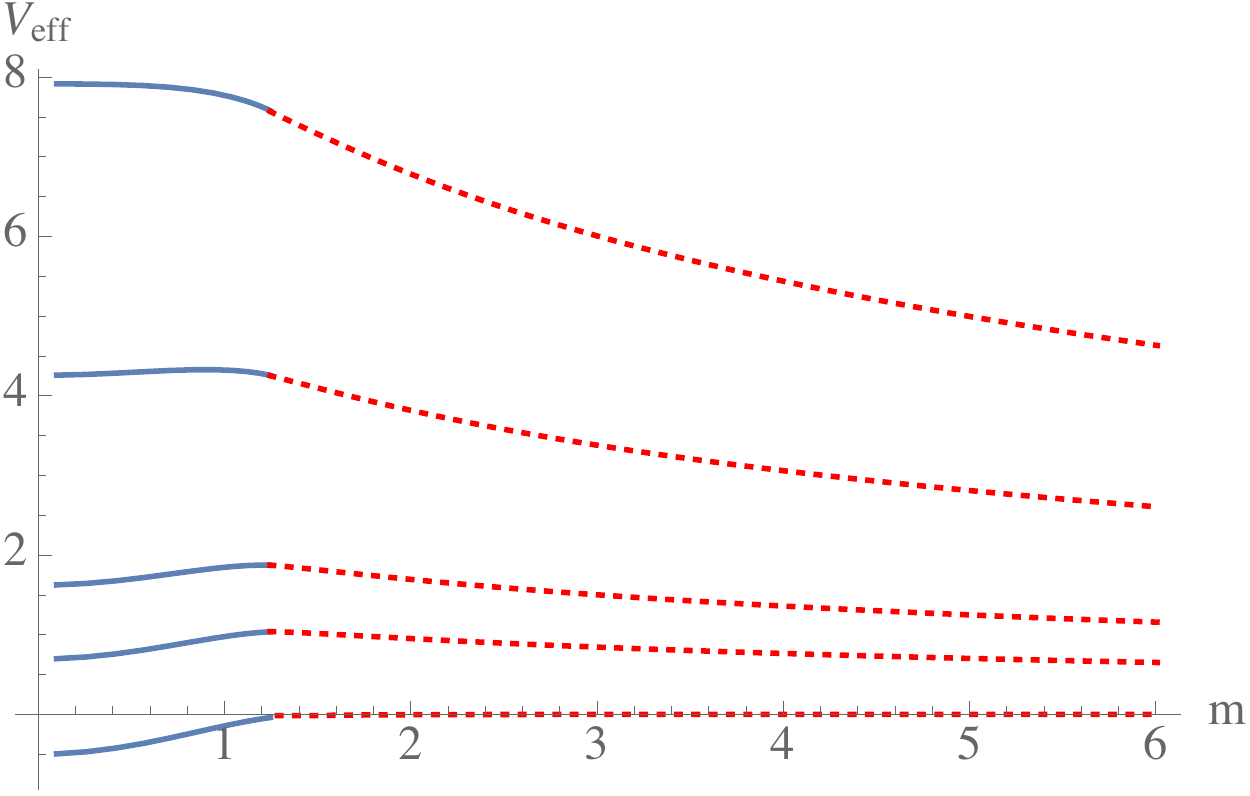}  \vspace{0.5cm}  }
  \caption{ The theory with $B < 4 w_H^2$ lies in a deconfined phase with D7 embeddings of the form shown in the top figure (here $B=w_H=1$). Heavy quarks are described by Minkowski the red dotted embeddings. Light quarks correspond to the solid  black hole embeddings. In the lower diagram we show the effective potential against $m$ without a NJL term - $B/w_H^2= 0, 0.75, 1, 1.5, 2$ from the bottom. }
            \label{fig:5}
\end{figure}

\begin{figure}[]
 \centering 
{
\includegraphics[width=6.5cm]{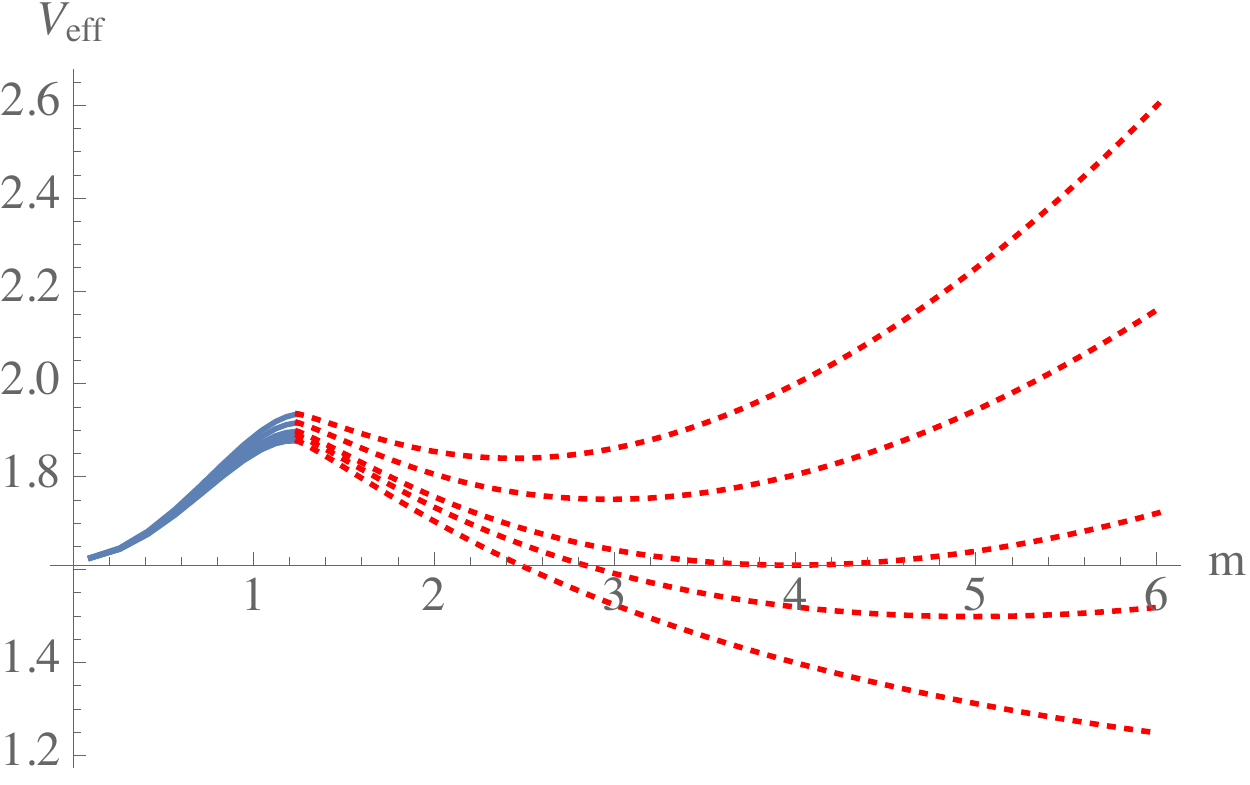} \vspace{0.5cm} \\ 
\includegraphics[width=6.5cm]{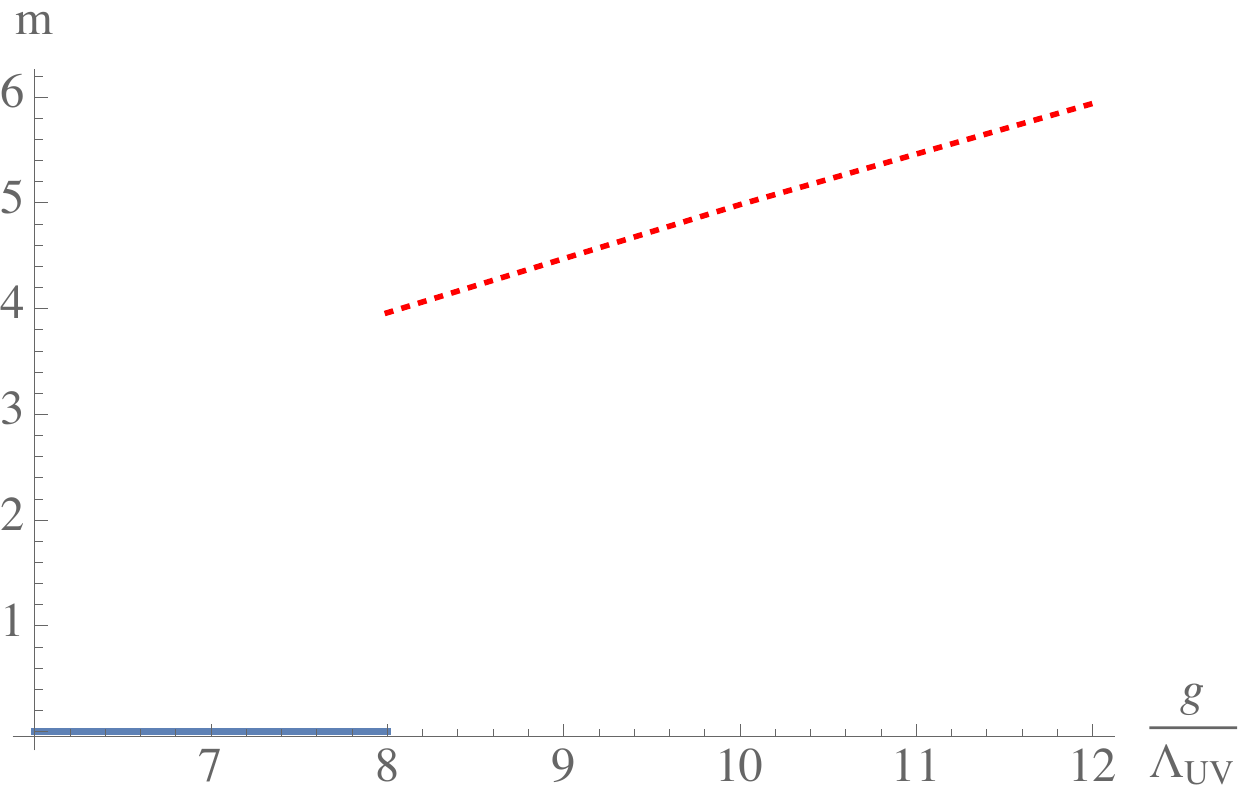}   }
  \caption{The D3/D7 model at finite temperature and magnetic field, $w_H=1$ and $B=1$, and with an NJL interaction at a UV cut off. Solid(red dotted) curves are for Minkowski(black hole) embeddings. \\
 {\bf Top:}  The effective potential versus $m$ for the finite $T$ and $B$ configurations with an NJL interaction and $\Lambda_{UV}=100$ - we show the cases $g/\Lambda_{UV} = 5, 6, 8, 10, 20$ from top to bottom. 
\\{\bf Bottom:} $m$ that minimizes the potential versus $g/\Lambda_{UV}$ showing the first order transition at  $g/U_{UV} \approx 8$. }
            \label{fig:6}
\end{figure}

The IR boundary condition remains as $L'(0)=0$ except the embedding can also potentially end on the black hole horizon. 
When $B > 4 w_H^2$ the embeddings for all $m$ are Minkowski and take the qualitative form of those in Fig \ref{fig:4}. The behaviour with an NJL interaction therefore follows the same generic behaviour of the previous section. When $B < 4 w_H^2$there is a phase transition \cite{TB} to a chirally symmetric phase where the embeddings for light quark masses lie on the horizon.  Here we will concentrate on this high temperature phase, setting $w_H=1$ and varying $B$.

Firstly we can set $B=0$. We can plot the effective potential derived from the embeddings' action (\ref{action}) against the quark mass (of course without an NJL term one should not minimize this potential - the mass is a fixed UV parameter but the form of the potential will show the behaviour with an NJL term present). In the lower graph of Fig \ref{fig:5} we show this potential as the lowest curve. The minimum lies at $m=0$ and the potential grows with $m$. The reason here that the action can be lower for embeddings that probe the deep IR is that the horizon truncates the space reducing the D7 world volume over which the action is computed. Note there is a discontinuity at $m=1.22$ where the embeddings move off the black hole. This marks the famous meson-melting phase transition \cite{Babington:2003vm,Mateos:2006nu} where the embedding switches at a first order transition from a black hole embedding to a Minkowski embedding. Were we to introduce an additional NJL term into the theory then, as we have seen, we would add an effective potential piece $\sim m^2$ which would further lift the potential at non-zero $m$. The minimum of the potential for all values of the NJL coupling remains zero. Given the base ${\cal N}=2$ quark theory does not generate chiral condensation with an NJL term it is no surprise that adding temperature does not change this situation. 

We can uncover some more interesting behaviour as we turn up a magnetic field. In the lower plot of Fig \ref{fig:5} we show the effective potential in the theory without an NJL term for various values of $B$. The effect of $B$ is to raise the level of the potential but more so in the IR than the UV. When $B \simeq 0.75 w_H^2$ the UV potential falls in value below the $m=0$ value. Above this value of $B$ an NJL operator can generate a chiral symmetry breaking vacuum. We show this in detail for one case $B = w_H^2=1$ and $\Lambda_{UV}=100$ in Fig \ref{fig:6}. For large values of the NJL coupling chiral symmetry breaking is observed but there is a first order phase transition to a chirally symmetric theory at $g/ \Lambda_{UV}^2 = 8$. Even at large $g$ the $m=0$ configuration is a metastable vacua. The first order transition is also a meson melting transition.

\section{\bf Goldstone Mode}

It is important to stress that all of the theories we have studied have no bare mass term. The mass is always dynamically generated by the NJL interaction. This means that all these configurations are associated with a Goldstone boson. If one considers the embeddings of Fig \ref{fig:3},\ref{fig:4} or \ref{fig:5} but now without an NJL term then each embedding corresponds to a different UV hard mass - naively there is a flat direction in the potential associated with rotating the embedding in the $\varphi$ direction of the metric. However, such a state is not a physical state in the theory because the asymptotic fall off of the perturbation from the vaccum goes as $m + c/\rho^2$ not as $c/\rho^2$. When though the configurations are viewed in the light of the NJL theory then $m \propto c$ and rotating the embedding in $\varphi$ is then just changing the phase of the condensate $c$ - they are now physical states. One can therefore see the flat direction in the potential (rotations in $\varphi$) that generate the Goldstone mode fluctuations. \bigskip

\section{\bf Summary}

We have introduced NJL operators into the D3/probe D7 system using the double trace prescription of Witten. In the basic system there are no embeddings of the D7 that satisfy both the IR and the new UV boundary conditions other than $L=0$, the flat embedding with no mass or condensate. To understand this we showed that the effective potential of the model without the presence of an NJL interaction is flat with the quark mass as a result of the ${\cal N}=2$ supersymmetry. This is distinct from the usual NJL model where the potential is unbounded to large $m$. We have argued that IR deformations of the D3/D7 system will lift the effective potential at small mass and so restore NJL like behaviour with dynamical mass generation above a critical NJL coupling. The precise critical coupling depends on the IR deformation. We have exhibited several examples. With a magnetic field present but an IR cut off, so it does not itself generate chiral symmetry breaking, we saw a second order transition at the NJL chiral transition. This model is very reminiscent of the original NJL model. In the case where the magnetic field is present with no IR cut off the B field itself triggers chiral symmetry breaking and we have seen that the NJL interaction then enhances the condensate smoothly as it is switched on and strengthened. Finite temperature usually disfavours chiral symmetry breaking and it indeed generates an effective potential that does not allow the NJL operator to generate condensation. We found an interesting example with both $T$ and $B$ present though. For small NJL interaction the temperature forbids the condensation but at larger values there is a first order transition to chiral symmetry breaking.  
We believe these  cases  provide examples of all possible generic behaviours with NJL interactions.

The overall picture is very satisfying and provides a new controlled example of chiral symmetry breaking/ quark condensate formation in a top down holographic model. The methodology may also be of interest for looking at extended technicolour \cite{Dimopoulos:1979es} type interactions in holographic models of technicolour dynamics such as \cite{Alho:2013dka}. The similar role of four fermion operators in holographic superconductors was explored in \cite{Faulkner:2010gj}

\noindent {\bf Acknowledgments: }The work of KK was supported by Basic Science Research Program through the National Research Foundation of Korea(NRF) funded by the Ministry of Science, ICT \& Future Planning(NRF-2014R1A1A1003220)  and the 2015 GIST Grant for the FARE Project (Further Advancement of Research and Education at GIST College). NE is supported by a STFC consolidated grant.  

\end{document}